\documentclass{revtex4}
\usepackage{graphicx}

\newcommand{\beq}{\begin{equation}}
\newcommand{\eeq}{\end{equation}}
\newcommand{\beqd}{\begin{displaymath}}
\newcommand{\eeqd}{\end{displaymath}}
\newcommand{\beqa}{\begin{eqnarray}}
\newcommand{\eeqa}{\end{eqnarray}}

\newcommand{\m}{\mu}

\newcommand{\comment}[1]{}

\newcommand{\dq}{\delta q}

\begin{document}

\title{Replica Symmetry Breaking Transitions and Off-Equilibrium Dynamics}

\author{Tommaso Rizzo$^{1,2}$}
\affiliation{
$^1$ IPCF-CNR, UOS Rome, Universit\`a "Sapienza", PIazzale A. Moro 2, \\
$^2$ Dip. Fisica, Universit\`a "Sapienza", Piazzale A. Moro 2, I-00185, Rome, Italy \\
I-00185, Rome, Italy}

\begin{abstract}
I consider branches of Replica-Symmetry-Breaking (RSB) solutions in Glassy systems that display a dynamical transition at a temperature $T_d$ characterized by a Mode-Coupling-Theory dynamical behavior. Below $T_d$ these branches of solutions are considered to be relevant to the complexity and to off-equilibrium dynamics.
Under general assumptions I argue that near $T_d$ it is not possible to stabilize the  one-step (1RSB) solution beyond the marginal point by making a full RSB (FRSB) ansatz. 
However, depending on the model, it may exist a temperature $T_*$ strictly lower than $T_d$ below which the 1RSB branch can be continued to a FRSB branch.
Such a temperature certainly exists for models that display the so-called Gardner transition and in this case $T_G<T_*<T_d$.
An analytical study in the context of the truncated model reveals that  the FRSB branch of solutions below $T_*$  is characterized by a two plateau structure and it ends where the first plateau disappears.
These general features are confirmed in the context of the Ising $p$-spin with $p=3$ by means of a numerical solution of the FRSB equations.
The results are discussed in connection with off-equilibrium dynamics within Cugliandolo-Kurchan theory.
In this context I assume that the RSB solution relevant for off-equilibrium dynamics is the 1RBS marginal solution in the whole range $(T_*,T_d)$  and it is the end-point of the FRSB branch for $T<T_*$.
Remarkably under these assumptions it can be argued that $T_*$ marks a qualitative change in off-equilibrium dynamics in the sense that the decay of various dynamical quantities changes from power-law to logarithmic.
\end{abstract}
\maketitle

\section{Introduction}

The connection between the replica method and dynamics is one of the most interesting feature of mean-field Spin-Glass (SG) models \cite{Kirkpatrick1987,Kirkpatrick1989,Crisanti92,Crisanti93,Cugliandolo93,Monasson95,Franz97}.
This connection is more striking in the context of the so-called one-step replica-symmetry-breaking (1RSB) models. 
Equilibrium dynamics in these models exhibits at some temperature $T_d$ a dynamical transition characterized by the fact that the spin-spin correlation function at different times no longer decays to the static equilibrium value but remains blocked at a higher value $q$. Notably the dynamical behavior approaching $T_d$ from above exhibits the same two-step relaxation predicted within mode-coupling-theory (MCT) \cite{Gotze09}.
Surprisingly this purely dynamical phenomenon can be captured within a simpler static replica computation where it corresponds to the abrupt appearance of a 1RSB solution with a Parisi breaking parameter $m=1$. This implies that both the value of $q$ and $T_d$ can be obtained by means of the replica method. More recently \cite{ParisiRizzo13} it has been realized that the replica method can be used to extract also the so-called parameter exponent $\lambda$, that controls the MCT exponents $a$ and $b$.
At temperatures lower than $T_d$ 1RSB systems are no longer able to reach equilibrium starting from a random configuration and exhibit aging. Quite remarkably the off-equilibrium aging regime for temperature $T<T_d$ has a structure that resembles the phenomenology of equilibrium MCT for temperature  $T>T_d$.
Much as in equilibrium the main observable is the spin-spin correlation defined as
\beq
C(\tau+t_w,t_w) \equiv {1 \over N} \sum_{i=1}^N \overline{\langle s_i(\tau+t_w)s_i(t_w)\rangle}\, ,
\eeq
where the square brackets are thermal averages and the overbar means disorder average.
According to the Cugliandolo-Kurchan (CK) scenario at large values of the waiting time $t_w$ the correlation has a two-step behavior as a function of $\tau$.
More precisely there is an initial relaxation towards a plateau value $q$ (similar to the $\beta$ regime  in structural glasses) followed by a second relaxation to zero on much larger times.  The first regime is called equilibrium regime because it turns out that the correlation and response functions obey the Fluctuation-Dissipation-Theorem (FDT). The second regime is called the aging regime and is characterized by the remarkable property that the response and correlation still obey FDT but with a lower effective temperature $T_{eff}=T/X$.
In the thermodynamic limit these systems never reach equilibrium and in particular one-time quantities like the energy approach at large times a limiting value different from the equilibrium one. 
Quite surprisingly it was found that in the spherical model the value of the the plateau $q$, the limiting value of the energy $E_{off}$ and the value of the FDT-violation parameter $X$ are the same that can be obtained considering a 1RSB solution of the replicated equilibrium theory with Parisi breaking parameter $m=X<1$ determined by the so-called {\it marginality condition}, {\it i.e.} requiring that the so-called replicon eigenvalue vanishes \cite{Almeida78}. It has been conjectured that the connection between off-equilibrium dynamics and the marginality condition holds for generic 1RSB models and positive evidences in favor of its validity was presented in \cite{Marinari94} although its origin remained somehow obscure.

The connection between RSB solutions and off-equilibrium dynamics has also been motivated in the context of  the Thouless-Anderson-Palmer (TAP) equations. At low temperatures there are many TAP solutions and the logarithm of their number (the so-called complexity) is $O(N)$, where $N$ is the system size. 
It can be argued \cite{Monasson95} that the complexity $\Sigma(f)$ of TAP solutions with a given free energy $f$ can be obtained from the free energy $\phi(\beta,m)$ of the 1RSB solution with breaking point $m$ by means of the following formulas:
\beq
\Sigma=\beta m^2 \partial_m \phi(\beta,m)\, , \ f=\partial_m[m \phi(\beta, m)]
\label{comprep}
\eeq
In the spherical model the marginal solution relevant for off-equilibrium dynamics is also the solution that corresponds to the maximal complexity. On the other hand in \cite{Montanari03} it was noticed that, at variance with the spherical $p$-SG model, in the case of Ising $p$-spin model the 1RSB solution that gives the maximum of the complexity as a function of $m$ does not coincide with the marginal solution. Furthermore the maximal solution has a negative replicon eigenvalue and therefore it is likely to be unphysical. A maximal complexity criterion was advocated in \cite{Montanari03} in order to determine the RSB solution relevant for off-equilibrium dynamics 
and it was claimed that in order to attain the states with maximal complexity the 1RSB branch of solutions has {\it always} to be continued to a full-RSB (FRSB) branch.  As consequence 1RSB aging as discussed by Cugliandolo and Kurchan \cite{Cugliandolo93}  only applies to the spherical model. However the claimed FRSB branch was not exhibited at any finite temperature, while an approximate 2RSB solution was computed at zero temperature.  Subsequent analytical studies on the complexity of TAP equations indicate that the 1RSB branch of solutions is not followed by a FRSB branch but rather by a branch that breaks the Becchi-Rouet-Stora-Tyutin (BRST) invariance \cite{Crisanti05} and these findings were later validated numerically in \cite{Tonosaki07}. 
At the present level of knowledge there is no evidence that the BRST-breaking states play any role in off-equilibrium dynamics. 
Assuming that the FRSB branch does not exists and that the BRST-breaking solutions are  irrelevant one could think that off-equilibrium dynamics is just associated to the marginal 1RSB solution. However this assumption may lead to the following paradox.
Many SG models, like the the Ising $p$-spin, exhibit at the so-called Gardner temperature $T_G$ a phase transition where the equilibrium RSB solution changes continuously from 1RSB to FRSB \cite{Gardner85}. It turns out that the static 1RSB equilibrium solution and the marginal 1RSB solution coincide at $T_G$ \cite{Montanari03,Crisanti05}  and therefore the marginality criterion would yield the rather absurd prediction that the system is not able to reach equilibrium at temperatures greater than $T_G$ while it would be able to do so at $T_G$.
This paradox can be avoided by means of the following results that will be presented in this paper: 

\begin{itemize}
\item Near the dynamical temperature of any SG system the 1RBS branch cannot be continued after the marginal point to a FRSB branch.

\item Depending on the model, it may exist a temperature $T_*<T_d$ below which the 1RSB branch of solutions can be continued to a FRSB branch. 

\item  The temperature $T_*$ {\it must} exist for models that display a Gardner transition and in this case $T_G<T_*$. In principle it can also exist for models where there is no Gardner transition.

\item At temperatures below $T_*$ the branch of FRSB solutions  displays some general features. Notably the end point of the FRSB branch has a higher value of the energy of the 1RSB marginal solution and it is  thus a natural candidate to yield the off-equilibrium energy and solve the off-equilibrium energy paradox at $T_G$. 
\end{itemize}
The above results concern essentially the existence and structure of particular branches of RSB solution. However the interesting question is their relevance to off-equilibrium dynamics.
In particular I will consider the following scenario:
\begin{itemize}
\item  Between $T_d$ and $T_*$ off-equilibrium dynamics displays 1RSB type of aging as described by CK. In particular the large time limit of the energy is given by the 1RSB marginal solution.

\item Below $T_*$ off-equilibrium dynamics is still of the CK type but with a continuous set of scales \cite{Cugliandolo95}. In this case the limiting value of the off-equilibrium energy and the function $X(q)$ are given by the end point of the FRSB branch.
\end{itemize}
I will not put to test this scenario by directly studying the dynamical equations, instead I will {\it assume} its validity and explore the implications.
The most interesting prediction is that the temperature $T_*$ marks a {\it qualitative} change in off-equilibrium dynamics. More precisely at $T_*$ the functional form of the long-time behavior of various off-equilibrium quantities changes from power-law to a much slower logarithmic decay. In a sense {\it the dynamical transition occurring at $T_*$ can be seen as the off-equilibrium analog of the so-called $A_3$ \cite{Gotze09} singularity within equilibrium MCT}.

The paper is organized as follows.
In section \ref{general} I will give a detailed presentation of the results and discuss them in connection with off-equilibrium dynamics.
The peculiar structure of the FRSB branch of solutions will be also described. I will quote some results that will be derived in sections \ref{onset} and \ref{structure-full}.
These results are essentially model-independent and will be indeed confirmed {\it a posteriori} by means of an explicit computation in the context of the Ising p-spin model presented in section \ref{Ipspin}.
In section \ref{numerics} I will present the outcome of off-equilibrium numerical simulations. 
Section \ref{conclusions} gives the conclusions.

\section{1RSB, FRSB and Off-Equilibrium Dynamics}
\label{general}

\subsection{Absence of the FRSB branch near the Dynamical Temperature}

In this section we will give a general argument to show that the 1RSB branch of solutions cannot be continued to a FRSB branch near $T_d$.
Let us start by recalling the properties of the phase diagram of the 1RSB solutions with Parisi breaking point $m$ in the $(T,m)$ plane for a generic model with a discontinuous transition. An instance of such a phase diagram in the case of the Ising $p$-spin is displayed in fig. (\ref{figTm}).
The dynamical transition temperature $T_d$ is characterized by the appearance of a 1RSB solution with $m=1$.
This solution is marginally stable because the so-called replicon eigenvalue vanishes. Actually the abrupt appearance of a solution leads to the vanishing of the so-called longitudinal eigenvalue but at $m=1$ the two eigenvalues are degenerate, see {\it e.g.} \cite{Franz11}. Precisely at $T_d$ the solution disappears as soon as $m<1$. At temperatures slightly lower than $T_d$ the solution can be analytically continued to values $m<1$. Furthermore the replicon eigenvalue is positive at $m=1$ and remains  stable for $m<1$ down to a value $m_G(T)$. The branch of solutions can be continued to even lower values of $m<m_G(T)$ down to the spinodal point $m_{spinodal}(T)$ where the solution disappears abruptly and correspondingly the longitudinal eigenvalue vanishes.
For $T<T_d$ the TAP complexity computed from these replica solutions according to the standard recipe (\ref{comprep}) attains a maximum as a function of $m$ at an intermediate value between $m_G(T)$ and $m_{spinodal}(T)$. This value is called $m_d(T)$ in \cite{Montanari03} because the maximal complexity criterion is advocated to select the 1RSB solution relevant for off-equilibrium dynamics. However since for $m<m_G(T)$ the branch of 1RSB solutions has a negative replicon $m_d(T)$ cannot actually have any physical meaning and in \cite{Montanari03} it is claimed that the true stable maximum $m_d(T)$ must be attained by continuing the 1RSB branch to a FRSB branch. Therefore the first question we want is to consider is the following: it is actually possible to stabilize the 1RSB branch of solutions for $m<m_G(T)$ by considering a FRSB ansatz? The answer is no, at least near $T_d$.
In order to understand why it is so we have to go back to some recent results concerning equilibrium MCT dynamics at $T_d$. 

The dynamical exponent $a$ and $b$ characterizing the $\beta$ and $\alpha$ regimes near $T_d$ are controlled by the so-called parameter exponent through the following relationship \cite{Gotze09}:
\beq
{\Gamma^2(1-a) \over \Gamma(1-2 a)}={\Gamma^2(1+b) \over \Gamma(1+2 b)}=\lambda \ .
\label{lambdadef}
\eeq
Therefore physical values of $\lambda$ are constrained between zero and one. The case $\lambda=1$ however is qualitatively different from the case $\lambda<1$. The latter describe a standard dynamical MCT transition characterized by well defined exponents $a$ and $b$, the former instead leads to $a=b=0$ and corresponds to a  different type of dynamical singularity (called $A_3$ in MCT literature) characterized by logarithmic decays instead of power-laws.  
Therefore since we are considering systems with the standard MCT phenomenology we will assume that $\lambda<1$.

In \cite{ParisiRizzo13} it has been shown that $\lambda$ can be computed from the replica method. One has to consider the expansion of the replicated Gibbs free energy near the 1RSB solution with $m=1$ at $T_d$ at third order.  The expansion in general has the following form \footnote{for the sake of simplicity we are retaining only the relevant cubic terms, see section \ref{onset} for the full expression}:
\beqa
G(\delta q) & = & {1 \over 2}\left( m_1 \sum_{ab} \delta q_{ab}^2+ m_2 \sum_{abc}\delta q_{ac}\delta q_{ab}+ m_3 \sum_{abcd}\delta_{ab}\delta q_{cd}\right)+
\nonumber
\\
& - & {1 \over 6} \left( w_1 \sum_{abc}\dq_{ab}\dq_{bc}\dq_{ca}+w_2 \sum_{ab}\dq_{ab}^3    \right)    
\label{GIBBS0}
\eeqa
and one has to determine the coefficients $w_1$ and $w_2$.
Once they are computed the parameter exponent is given by the following formula:
\beq
\lambda={w_2 \over w_1} \ .
\eeq
From the above discussion it follows that the ratio $w_2/w_1$ must be definitively smaller than one at $T_d$.
Now we turn to the replica problem and consider the possibility of stabilizing the 1RSB solution with breaking point $m$ smaller than $m_G(T)$ by considering Parisi function $q(x)$ that exhibits FRSB in the region $x>m$. In order to do so I will show that one should consider a $q(x)$ with a continuous part localized near the point $x=w_2/w_1$, where the two coefficients are computed with respect to the 1RSB solution with $m=m_G(T)$. This result is a generalization of earlier results and its detailed derivation will be postponed to section \ref{onset}. To complete the argument we note that at $T=T_d$ we have $m_G=1$ and $w_2/w_1=\lambda<1$. Since $m_G$, $w_1$ and $w_2$ are continuous functions of the temperature it follows that for temperatures smaller than but close to $T_d$ we will still have $w_2/w_1<m_G$ and therefore we cannot continue the 1RSB branch to a FRSB branch because we should put the continuous part of the $q(x)$ at values of $x$ smaller than the breaking point $m_G$. 

\subsection{The $T_*$ transition Temperature and the structure of the FRSB solution below $T_*$}

The above argument guarantees that near the dynamical temperature no FRSB branch of solutions exists after the 1RSB marginal point $m_G(T)$. The argument is purely topological and it does not necessarily holds at all temperatures. In particular it may exist a temperature $T_*$ where the ratio $w_2/w_1$ computed on the marginal 1RSB solution is equal to $m_G$. As we will see in the following for temperatures $T<T_*$  the 1RSB branch {\it can} actually be continued to values of $m$ smaller than $m_G(T)$  by considering a FRSB ansatz.

The existence of the transition temperature $T_*$ depends on the model, however one can argue that {\it $T_*$ must exists for models that present a Gardner transition at some temperature $T_G<T_d$}.
Indeed the Gardner temperature by definition marks the position where the static replica solution changes continuously from 1RSB to FRSB, by developing a continuous part for $x>m_s(T)$, where $m_s(T)$ is the breaking point of the equilibrium 1RSB solution \cite{Gardner85}. The second-order nature of the transition implies that  $m_s(T_G)=m_G(T_G)$ (leading to the energy paradox discussed in the introduction) and the above argument implies that at $T_G$ we must have $w_2/w_1>m_G$. It follows that since $w_2/w_1>m_G$ at $T_G$ and  $w_2/w_1<m_G$ at $T_d$ it must exist an intermediate temperature $T_*$ where $w_2/w_1=m_G$.

The FRSB branch of solutions for $m<m_G(T)$ and $T<T_*$ will be studied in section \ref{structure-full} in the context of the so-called truncated model introduced by Parisi \cite{Parisi79}. The qualitative features of the solutions are likely to be general and indeed will be recovered also  in the Ising $p$-spin model that will be studied in section \ref{Ipspin}. 
The function $q(x)$ for $m<m_G(T)$ has a discontinuity at the breaking point $m$ followed by a continuous part according to the following structure:
\beqa
q(x)=q_m &  \mathrm{for} & x<m
\nonumber
\\
q(x)=q_m=q(x_p)  &  \mathrm{for} & m<x< x_p  
\nonumber
\\
q(x)  &  \mathrm{for} & x_p<x<x_P 
\nonumber
\\
q(x)=q_1=q(x_P)  &  \mathrm{for} & x_P <x<1
\nonumber
\eeqa
Therefore the continuous part of the FRSB solution is characterized by a plateau between $m$ and $x_p$, an increasing part between $x_p$ and $x_P$ and a second plateau between $x_P$ and $1$. 
Note that even for  $T<T_*$ the solution is 1RSB for $m>m_G(T)$. Decreasing $m$ below $m_G$ the two plateaus develop in a continuous fashion with a small continuous region between them concentrated near the point $x=w_2/w_1$. Decreasing $m$, the difference in height of the two plateaus increases while the length of first plateau decreases until it shrinks to zero at some value $m_{end}(T)$. This point is the end-point of the FRSB branch because analytical  continuation to smaller values of $m$ would require a plateau with negative length.

We note that for $T>T_*$ the end point of the 1RSB branch of solutions is identified by the marginality condition. This condition cannot work for $T<T_*$ because {\it all solutions for $m_{end}(T)<m<m_G(T)$ are marginal} due to FRSB (see \cite{DeDominicis98} an references therein). Therefore it is rather satisfactory to have an alternative precise characterization of the end point as the point where the the first plateau disappears.

\subsection{Off-Equilibrium Dynamics}
\label{suboff}

In the following I will discuss off-equilibrium dynamics in the light of the previous results.
I will not study directly off-equilibrium dynamics but rather work under the assumption that the connection with RSB observed in the spherical model holds in general. 
More precisely I will assume that: i) off-equilibrium dynamics is described by CK theory with a scale-dependent FDT function $X(q)$ that can be obtained from a replica computation, ii) the RSB solution relevant for off-equilibrium dynamics is the 1RBS marginal solution in the range $(T_*,T_d)$  and iii) the RSB solution relevant for off-equilibrium dynamics is the end-point of the FRSB branch for $T<T_*$.
A natural consequence of these assumptions is that the long-time limit of the off-equilibrium energy is given by the energy of the corresponding RSB solutions, leading to the solution of the energy paradox implied by the marginality condition at $T_G$.

The above assumptions has further interesting implications on off-equilibrium dynamics, namely a qualitative change at $T_*$.
Off-equilibrium dynamics  in 1RSB systems displays within the CK scenario a considerable degree of similarity with the glass transition singularity of equilibrium MCT.
In particular it turns out that the initial relaxation  of the correlation towards the plateau value $q$ is described by a power-law decay similarly to the $\beta$ regime in structural glasses \cite{Bouchaud98}:
\beq
C(\tau+t_w,t_w) \approx q+ {c_a \over \tau^a} \ ,
\eeq
while the early stage of the subsequent decay from the plateau are described by a different exponent $b$
\beq
C(\tau+t_w,t_w) \approx q-{c_{b} \left({\tau \over {\mathcal T}_w}\right)^b}\ ,
\eeq 
where ${\mathcal T}_w$ is a time scale that depends on $t_w$.
In the context of the spherical model it was found \cite{Cugliandolo96} that the two exponents $a$ and $b$ obey the following relationship that generalizes  eq. (\ref{lambdadef}) of MCT:
\beq
{\Gamma^2(1-a) \over \Gamma(1-2 a)}=X{\Gamma^2(1+b) \over \Gamma(1+2 b)}=\lambda \ .
\label{lambdaoff}
\eeq
where the off-equilibrium parameter exponent $\lambda$ can be computed from the model-dependent spherical Hamiltonian and $X$ is the FDT violation ratio.

Recently the connection between dynamics and replicas has been studied in the context of equilibrium theories of glassy systems \cite{ParisiRizzo13} and also in off-equilibrium situations \cite{Caltagirone13} for some SG models. Similar arguments, to be presented elsewhere, can be used also in the context of discontinuous SG in order to study the connection between RSB a off-equilibrium dynamics. In this context one can show that, {\it if} a 1RSB solution is actually relevant for off-equilibrium dynamics, then it must satisfy the marginality condition. Furthermore it can be argued that eq. (\ref{lambdaoff})  holds as well,  with the parameter exponent $\lambda$ given by the ratio $w_2/w_1$ computed expanding around the marginal 1RSB solution.
The last result has important implications on off-equilibrium dynamics at $T_*$. Indeed the presence of the factor $X$ in the second term of eq. (\ref{lambdaoff}) implies that the effective parameter exponent is actually $\lambda_{eff} \equiv \lambda/X$. This determines a {\it second} condition, besides the marginal one, on the 1RSB solution relevant to off-equilibrium dynamics, that is $\lambda_{eff} \leq 1$.
We see that at $T=T_*$ we have $\lambda_{eff}=w_2/(w_1\,m)=1$ and therefore  the 1RSB marginal solution must be abandoned below $T_*$ because it cannot describe consistently the decay from the plateau value.
Furthermore the dynamical exponent $b$ vanishes at $T_*$ meaning that the decay from the plateau is slower than a power law. This is similar to what happens at the so-called $A_3$ singularity in equilibrium MCT \cite{Gotze09}. This singularity is indeed characterized by $\lambda=1$ and as a consequence the equilibrium decay of various quantities changes from power-law to logarithmic \cite{Gotze89}. Summarizing {\it the dynamical transition occurring at $T_*$ is the off-equilibrium analog of the $A_3$ singularity}.

It is well known that the direct observation of the exponents $a$ and $b$ from data at finite $t_w$ is not easy. It is usually easier to work with off-equilibrium one-time quantities, say the energy.
Unfortunately the theory of off-equilibrium dynamics in mean-field spin-glass models is still incomplete, in the sense that we are not able to characterize the off-equilibrium behavior of one-time quantities in 1RSB systems. Observations in the spherical \cite{Cugliandolo93,Franz95,Kim01,Lefevre06} and in the Ising $p$-spin \cite{Montanari04} models suggest that the decay is power-law but how to compute the actual exponents is at present unknown. On the other hand one can imagine that this exponent is somehow related to the exponents $a$ and $b$, similarly to the case of continuous spin-glass models \cite{Caltagirone13}.  Then one would expect that $T_*$ should also corresponds to the vanishing of the energy exponent and that the decay changes from power-law to logarithmic at and below $T_*$.

\section{The onset of Full-Replica-Symmetry Breaking}
\label{onset}

In this section we derive one of the general results that we have used to argue that near $T_d$ there is no FRSB branch.
We will show that an unstable 1RSB solution may be stabilized by means of FRSB ansatz provided the ratio $w_2/w_1$ is larger than the breaking point $m$ of the 1RSB solution.
This is essentially a generalization of the result obtained originally by Kanter, Gross and Sompolinsky in the context of the Potts SG \cite{Gross85}.
The problem is essentially equivalent to a RS problem with $n$ replicas, where $n$ is equal to the breaking parameter $m$ of the 1RSB solution.
Therefore we work in the general case where the order parameter is a replicated matrix $q_{ab}$ of size $n \times n$ and we consider its power series expansion near the Replica-Symmetric solution: $q_{ab}=q+\delta q_{ab}$. 
The replicated Gibbs free energy of the block reads:
\beqa
G(\delta q) & = & {1 \over 2}\left( m_1 \sum_{ab} \delta q_{ab}^2+ m_2 \sum_{abc}\delta q_{ac}\delta q_{ab}+ m_3 \sum_{abcd}\delta_{ab}\delta q_{cd}\right)+
\nonumber
\\
& - & {1 \over 6} \left( w_1 \sum_{abc}\dq_{ab}\dq_{bc}\dq_{ca}+w_2 \sum_{ab}\dq_{ab}^3+    \right.    
\nonumber
\\
& +& w_3 \sum_{abc}\dq_{ab}^2\dq_{ac}+w_4 \sum_{abcd}\dq_{ab}^2\dq_{cd}+w_5 \sum_{abcd}\dq_{ab}\dq_{ac}\dq_{bd}+
\nonumber
\\
& + & \left. w_6 \sum_{abcd}\dq_{ab}\dq_{ac}\dq_{ad}+w_7 \sum_{abcde}\dq_{ac}\dq_{bc}\dq_{de}+w_8 \sum_{abcdef}\dq_{ab}\dq_{cd}\dq_{ef}   \right)                    \, ,
\label{GIBBS}
\eeqa
The quantity $\delta q_{ab}$ is determined by the conditions
\beq
{\partial G \over \partial \delta q_{ab}}=0
\label{DG}
\eeq
We work under the assumption that the solution with $\delta q_{ab}=0$ is slightly unstable meaning that the replicon eigenvalue (which is given precisely by $m_1$ \cite{Temesvari02}) is {\it small and negative}. The derivative of the replicated Gibbs free energy with respect to $\delta q_{ab}$ will contain many terms, however it can be checked straightforwardly that the only three therms that depend explicitly on {\it both} indexes $a$ and $b$ are:
\beq
{\partial G \over \partial \delta q_{ab}}=0=2 m_1 \delta q_{ab}+w_1 (\delta q)^2_{ab}+w_2 \delta q^2_{ab}+\dots
\eeq
where the dots represent term that depend explicitly on only one between the indexes $a$ or $b$ ({\it e.g.} $m_2 \sum_c \delta q_{ac}$ )  or do not depend at all on $a$ and $b$ ({\it e.g.} $m_3\sum_{cd}\delta q_{cd}$).

Now we make the Parisi ansatz on the matrix $\delta q_{ab}$ parameterizing it through the function $\delta q(x)$ where $n<x<1$ and we plug the ansatz into equation (\ref{DG}). Due to the nature of Parisi ansatz any combination of $\delta q_{ab}$ that depends on a single index ({\it e.g.} $m_2 \sum_c \delta q_{ac}$ ) is independent of the index $a$ (this property is called replica equivalence). As a consequence the only terms that depend explicitly on $x$ in the equations are precisely the terms that we have selected above.
This means that the equation of state can be rewritten as:
\beq
0=-2 m_1 \delta q(x)+  w_1 \left(-2 \overline{\delta q}\, \delta q(x)-n \, \delta q(x)^2-\int_n^x(\delta q(x)-\delta q(y))^2 dy\right)+w_2 \delta q(x)^2+ C 
\label{Pareq}
\eeq
where $\overline{\delta q} \equiv \int_{n}^1 \delta q(x)dx$ and $C$ is a constant that depend on the function $\delta q(x)$ and on all the remaining $m$'s and $w$'s but that does not depend explicitly on $x$.
Following Parisi \cite{Parisi80} we derive the above equation with respect to $x$, we divide by $\delta q(x)$ and we perform another derivative with respect to $x$, obtaining:
\beq
(w_1 \, x-w_2)\delta \dot{q}(x)=0\ .
\eeq
The above equation means that we can have $\dot{q}(x) \neq 0$  {\it i.e.} FRSB only in a small $O(m_1)$ region around the point $x=w_2/w_1$ and from this it follows that if $w_2/w_1<n$ we cannot have any FRSB. 

The behavior of $\delta q(x)$ in the small $O(m_1)$ region near $x_1=w_2/w_1$ ({\it e.g.} the slope $\delta \dot{q}(x_1)$) is controlled by the quartic terms not shown in equation (\ref{GIBBS}).
On the other hand while $\delta q(x)$ is $O(m_1)$ and therefore the terms written explicitly in eq. (\ref{Pareq}) are $O(m_1^2)$ the constant term contains term proportional to $m_2$ and $m_3$ that would be $O(m_1)$ unless the following condition holds:
\beq
\sum_c \delta q_{ac}=\overline{\delta q}=\int_n^1 \delta q(x)dx=O(m_1^2) \ .
\eeq
Technically this can be also seen as a manifestation of the regular nature of the longitudinal eigenvalue.
The above quantity depends explicitly on all the $m$'s and all the $w$'s, instead the function $\delta q(x)$ at leading order depend solely on $m_1$, $w_1$ and $w_2$. The function is defined indeed by the height of the two plateaus separated by the small region near $x_1=w_2/w_1$ where $\delta q(x)$ is continuous.
Considering the difference between eq. (\ref{Pareq}) evaluated at $x=n$ and at $x=1$ we can remove the constant $C$ and  obtain an equation for $\delta q(1)$ and $\delta q(n)$:
\beq
0=(-2 m_1 -2 w_1 \overline{\delta q} )(\delta q(1)-\delta q(n))+   w_1(n-x_1)(\delta q(1)-\delta q(n))^2+(w_2-n w_1) (\delta q(1)^2-\delta q(n)^2)\ 
\eeq
On the other hand the condition $\overline{\delta q}=O(m_1^2)$ leads to a second equation:
\beq
\delta q(1)(1-x_1)+\delta q(n)(x_1-n)=O(m_1^2)\ .
\eeq
and the two equations fix the values of the two plateaus:
\beq
\delta q(n)={m_1 \over  w_1 (x_1-n)}+O(m_1^2)\, , \ \  \delta q(1)={-m_1 \over  w_1 (1-x_1)}+O(m_1^2)\, .
\label{plateaus}
\eeq
Note that $\delta q(n)$ is negative while $\delta q(1)$ is positive as it should.

In order to understand why the original 1RSB problem is essentially equivalent to the RS problem considered in this section one can use the following arguments. 
For models where the 1RSB ansatz is such that $q_0=0$ the different blocks of size $x \times x$ are uncorrelated, therefore it is evident that the action within each block is given precisely by the Gibbs free energy (\ref{GIBBS}) with $n$ equal to the breaking point $x$. 
In the case where $q_0\neq 0$ the actual Gibbs free energy will contain also a correlations between $\delta q_{ab}$ within different blocks. However in general these terms will produce regular correlations and one can argue that at order $O(m_1)$ the function $\delta q(x)$ inside each block will be given by the same expression above.

We note that the same results (\ref{plateaus}) for $\delta q(1)$ and $\delta q(n)$ together with the condition $x=w_2/w_1$ would be obtained considering considering a 1RSB $\delta q(x)$ and extremizing with respect to the breaking point $x$.  

\section{The Structure of the FRSB branch}
\label{structure-full}

In this section we will study the FRSB branch of solutions in the Ising $p$-spin model with $p=2+\epsilon$ with $\epsilon \ll 1$. In the case $p=2$ this is the Sherrington-Kirkpatrick (SK)  Model and near the critical temperature the FRSB solution can be obtained considering the so-called truncated model \cite{Parisi80}. As recognized originally by Kirpatrick and Thirumalay \cite{Kirkpatrick87} the advantage of the $2+\epsilon$ limit is that it is a model that has a weakly discontinuous transition that can be studied perturbatively. The region of the dynamical transition occurs at a distance $\epsilon \ln \epsilon$ from the SK transition temperature see {\it e.g.} section III.a in \cite{Ferrari12} where the parameter $w_2/w_1$ at $T_d$ is computed.

In the following we will focus on a region of the parameter space where the solution can be see as a perturbation of the solution with $q=0$. One can argue that the equation for Parisi's $q(x)$ for the problem is the same of the truncated model plus a term that vanishes for $\epsilon=0$:
\beq
2 \, (\tau-\overline{q}) \, q(x)+y q^3(x)-\int_0^x(q(x)-q(y))^2dy+(q(x)-q^{1-\epsilon}(x))=0
\label{tru1}
\eeq
In order to study possible FRSB solutions of the above equation, following Parisi \cite{Parisi80} we differentiate the above equation with respect to $x$ and we divide the result by $\dot{q}(x)$ obtaining:
\beq
2 \, (\tau-\overline{q})  +3 y q^2(x)-2\int_0^x(q(x)-q(y))dy+(1-(1-\epsilon)q^{-\epsilon}(x))=0 \ .
\label{tru2}
\eeq
Differentiating once again we obtain the condition that the continuous part of $q(x)$ (where $\dot{q}(x)\neq 0$) obeys the following equation
\beq
x(q)=3yq+{\epsilon (1-\epsilon) \over 2 q^{1+\epsilon}}\ .
\label{xtru}
\eeq
The function $x(q)$ for positive values of $q$ has a minimum different from zero for $\epsilon>0$ located at $x_{min}=\sqrt{6 \, y \, \epsilon}$. As a consequence the inverse function $q(x)$ can take two possible $q_+(x)>q_-(x)$. Both $q_+(x)$ and $q_-(x)$ are defined only for $x>x_{min}$. Near $x_{min}$ both approach the value $q_{min}=\sqrt{\epsilon /(6 \, y)}$ with a square-root singularity. The physical solution is the increasing one, that is $q_+(x)$.
We consider a FRSB solution parameterized by the three parameters $m$, $q_m$ and $q_1$ according to:
\beqa
q(x)=0 &  \mathrm{for} & x<m
\nonumber
\\
q(x)=q_m  &  \mathrm{for} & m<x< x_p \equiv x(q_m) 
\nonumber
\\
q(x)=q_+(x)  &  \mathrm{for} & x_p<x<x_P \equiv x(q_1)
\nonumber
\\
q(x)=q_1  &  \mathrm{for} & x(q_1)<x<1
\nonumber
\eeqa
Now evaluating eq. (\ref{tru1}) in $x_p$ divided by $q_m$ and subtracting eq. (\ref{tru2}) in $x_p$ we obtain the following equation for $q_m$ and $m$:
\beq
2 y q_m=m-{\epsilon \over q_m^{1-\epsilon}}  ,
\label{q0n}
\eeq
combining it with of eq. (\ref{xtru}) we can obtain:
\beq
x_p-m=y q_m-{\epsilon \over 2 q_m}\, +O(\epsilon\, q_m)
\eeq
the quantity $x_p-m$ is the size of the first plateau and it must be positive by definition. The first important thing that we note from the above expression is that it is {\it negative} when evaluated at the lowest possible value $q_m=q_{min}=\sqrt{6 \, y \, \epsilon}$ (where the function $q_+(x)$ has a the square-root singularity). This means that it is not possible to find a solution such that $q_m=q_{min}$. The lowest possible value of $m$ for which a FRSB solution can be obtained is thus the one in which the size of the first plateau is zero ($x_p=m$), which is given by:
\beq
q_m \simeq \sqrt{\epsilon \over 2 y}\ ,\ m_{end}\simeq 2 \sqrt{2 \epsilon \, y}\ ,
\eeq 
For $m>m_{end}$ the value of $q_m$ (and thus of $x_p$) is determined by
eq. (\ref{q0n}). In order to complete the characterization of the solution and determine $q_1$ we go back to eq. (\ref{tru1}) evaluated in $x_p$ and we divide it by $q_m$ obtaining:
\beq
2(\tau-\overline{q})+y q_m^2-q_m m+(1-q_m^{-\epsilon})=0
\label{eax0}
\eeq 
now $\overline{q}$ can be expressed in terms of $q_m$ and $q_1$ by means of the function $x(q)$ defined in (\ref{xtru}), the result is:
\beq
\overline{q}=q_1-m q_m-{3 y \over 2}(q_1^2-q_m^2)-{\epsilon \ln (q_1/q_m) \over 2}
\eeq
the above expression can be plugged into eq. (\ref{eax0}) yielding an exact equation between expressing $q_1$ in terms of $\tau$, $\epsilon$, $m$ and $q_m$. Eliminating $q_m$ by means of eq. (\ref{q0n}) we finally obtain:
\beq
2 \tau-2 q_1+3 y q_1^2=-\epsilon+O(\epsilon^2 \ln^2 \epsilon)
\eeq
For $\epsilon=0$ this reduces to the equation for $q_1$ in the truncated model as obtained originally in \cite{Parisi80}. Note that the leading  order correction to $q_1$ is $O(\epsilon)$ and it is independent of $m$ and that a small non-zero value of $\epsilon$ induces a regular $O(\epsilon)$ deviation on the $q(x)$ except in the region of small $x=O(\sqrt{\epsilon})$ where it produces an $O(\sqrt{\epsilon})$ deviation.

\section{RSB solutions in the fully connected Ising $p$-spin}
\label{Ipspin}

In this section we investigate the phase diagram of the fully connected Ising $p$-spin.
In the case $p=3$ we confirm the existence of a temperature $T_*$ between $T_d$ and $T_G$.
The solution below $T_*$ is studied solving numerically the FRSB equations. 
The fully-connected Ising $p$-spin is defined by the following Hamiltonian:
\beq
H=-\sum_{i_1<\dots<i_p}J_{i_1\dots i_p}s_{i_1}\dots s_{i_p}
\eeq
where the quenched random couplings $J$  have zero mean and variance $\overline{J^2}=p!/(2 N^{p-1})$.
By making the Parisi ansatz the free energy reads \cite{Gardner85}:
\beq
\beta \Phi=-{\beta^2 \over 4}\left[ 1-\int_m^1 q^p(x)dx+2\int_m^1 \lambda(x)q(x)dx-2 \lambda(1) \right]-{1 \over m}\ln \int_{-\infty}^{\infty}{1 \over \sqrt{2 \pi \lambda(n)}}\exp\left[ -{y^2\over 2 \lambda(m)}+\beta m f(m,y)\right]
\label{fvar}
\eeq
The above expression has to be extremized with respect to the Parisi functions $q(x)$ and $\lambda(x)$.
We recall that the parameter $m$ is the breaking point of the solution such that for $x<m$ we have $q(x)=\lambda(x)=0$.
The function $f(x,y)$ obeys the Parisi equation:
\beq
\dot{f}=-{\dot{\lambda} \over 2}\left( f''+\beta\, x (f')^2\right)
\eeq
with initial condition
\beq
f(1,y)={1 \over \beta}\ln 2 \cosh \beta y
\eeq
where dots are $x$-derivatives and primes are $y$-derivatives.
The variational equations for the free energy can be obtained using Lagrange multipliers \cite{Sommers84} and read:
\beq
\lambda(x)=p q^{p-1}(x)/2\, ;
\eeq
\beq 
q(x)=\int_{-\infty}^{\infty}P(x,y)\m^2(x,y)dy
\label{qvar}
\eeq
where $\m(x,y) \equiv f'(x,y)$ and
\beq
\dot{\m}=-{\dot{\lambda} \over 2}\left( \m''+2\, \beta\, x\, \m\,\m'\right) \ .
\label{eqmu}
\eeq
The function $\m(x,y)$ is usually called $m(x,y)$ in the literature but we renamed it to avoid confusion  with the breaking point value $m$. The auxiliary function $P(x,y)$ obeys:
\beq
\dot{P}={\dot{\lambda} \over 2}\left( P''-2\, \beta\, x\, (P\,\m)'\right)
\label{eqP}
\eeq
with initial condition at $x=m$:
\beq
P(m,y)= c \, \exp\left[ -{y^2\over 2 \lambda(m)}+\beta m f(m,y)\right]
\eeq
where $c$ is a normalization constant ensuring that $\int P(m,y)dy=1$.
Other equations can be obtained by repeated differentiation of the variational equations with respect to $x$ in the FRSB region. This is simplified by the use of the following Sommers identity \cite{Sommers85}:
\beq
{d \over dx}\int dy P\,g =\int dy P \, \Omega g 
\eeq
where $g(x,y)$ is any function and $\Omega$ is the following operator:
\beq
\Omega={\partial \over \partial x}+{\dot{\lambda} \over 2}\left( {\partial^2 \over \partial y^2} +2 \beta x \m(x,y){\partial \over \partial y} \right) \ .
\eeq
Deriving eq. (\ref{qvar}) and dividing by $\dot{q}(x)$ we obtain:
\beq
{2 q^{2-p}(x)\over p(p-1)}=\int dy P (\m')^2 \ .
\label{first}
\eeq
Repeating the process once again we obtain:
\beq
{4 (2-p)q^{3-2p} \over p^2(p-1)^2}=\int dy \, P(\m'')^2 - 2 \beta x \int P (\m')^3
\eeq
which can be rewritten as:
\beq
x={{ 4 (p-2)q^{3-2p}(x) \over p^2(p-1)^2}+\int P(\m'')^2 \over 2 \beta \int P(\m')^3}
\label{second}
\eeq
Equations (\ref{first}) and (\ref{second}) hold in the continuous region of the FRSB solution and in the general they are not satisfied by a 1RSB solution. However they must be satisfied at the point where the 1RSB branch can be continued to the FRSB branch, consistently one can check that the condition  (\ref{first}) evaluated on a 1RSB solution is precisely the marginality condition given in \cite{Gardner85}. Similarly it follows that near the marginal solution the continuous part of the FRSB solution is concentrated near a value of $x$ given by (\ref{second}). Therefore the r.h.s. of the eq. (\ref{second}) must be equal to the ratio $w_2/w_1$ according to the results in section (\ref{onset}) and indeed eq. (\ref{second}) agrees with the computation of $w_2/w_1$ at $T_d$ given by eqs. (50-52) in \cite{Ferrari12}.
Following \cite{Sommers85} one could perform another derivative of the equation in order to compute the value of $\dot{q}(x)$ at the breaking point.

\begin{figure}[htb]
\begin{center}
\includegraphics[width=10cm]{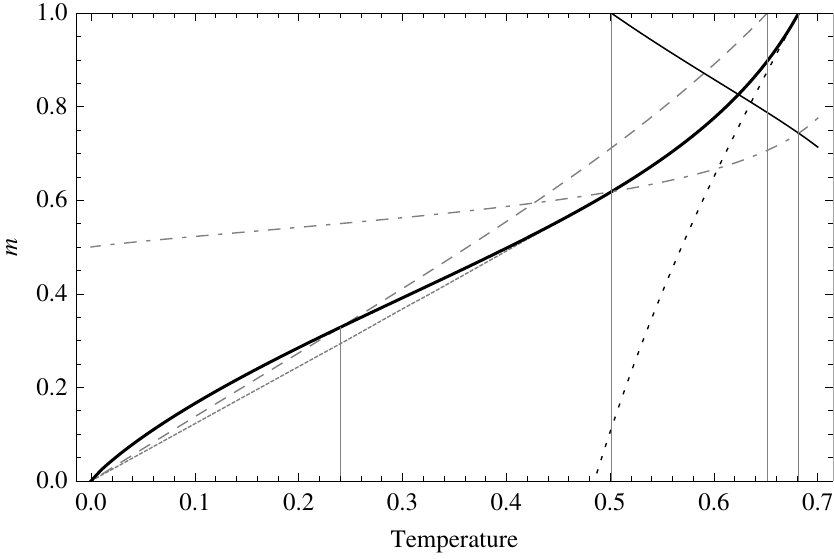}
%\put(-270,80){\rotatebox{90}{\it m}}
%\put(-130,-10){\it T}
\caption{Phase diagram in the $(T,m)$ plane for the $p=3$ Ising Spin-Glass. The thin vertical lines represent the temperatures: $T_G=0.24026$, $T_*=.501227$ $T_s=.651385$ and $T_d=.681598$. Dotted line: spinodal line $m_{spinodal}(T)$ of the 1RSB solution. Thick black line: marginal line $m_G(T)$ of the 1RSB solution. Dashed line: static 1RSB line $m_s(T)$, this coincides with the equilibrium solution between $T_s$ and $T_G$. Dashed-Dotted line: parameter exponent $\lambda(T)$ of the marginal 1RSB solution. Thin solid line: effective parameter exponent $\lambda_{eff}\equiv \lambda(T)/m_G(T)$. Solid gray line: breaking point $m_{end}(T)$ of the end-point solution of the FRSB branch.}
\label{figTm}
\end{center}\end{figure}

In figure (\ref{figTm}) we present the phase diagram in the $(T,m)$ plane of the case $p=3$.
The dotted line is the spinodal line $m_{spinodal}(T)$ of the 1RSB solution.  On the left of this line the 1RSB variational equations admit two solutions with $q>0$, besides the paramagnetic one $q=0$. For our purposes only the one with a larger value of $q$ is important and we will be referring to it in the following discussion. The two solutions merge on the spinodal line and disappear with a square root singularity leading to the vanishing of the longitudinal eigenvalue.

The thick black line is the marginal line $m_G(T)$  of the 1RSB solution where the replicon eigenvalue vanishes according to Gardner \cite{Gardner85} ore equivalently where the 1RSB solution satisfies eq. (\ref{first}).  
On the right of this line the 1RSB solution has negative replicon and therefore the whole region between the $m_{spinodal}(T)$ and $m_G(T)$ is unphysical. Note that the two lines crosses for $m=1$ and $T=T_d=.681598$ and this is consistent with the fact that at $m=1$ the replicon and longitudinal eigenvalues are degenerate leading to a non-trivial critical behavior at $T_d$ \cite{Franz11}. 

The dashed line corresponds to the breaking point $m_s(T)$ of the 1RSB solution that extremizes the free-energy (\ref{fvar}) as a function of $m$.
This solution is the equilibrium one in the range of temperatures between the static temperature $T_s=.651385$ and the Gardner temperature $T_G=0.24026$. The static temperature is identified by the condition $m_s(T_s)=1$ while the Gardner temperature is where the marginal line and the static line crosses: $m_G(T_G)=m_s(T_G)$. As shown by Gardner, for lower temperatures the static solution has a continuous FRSB structure for values of $x$ larger than the breaking point $m$.

The dashed-dotted line is the ratio $\lambda(T) \equiv w_2/w_1$ of the marginal 1RSB solution as a function of the temperature and it is given by the r.h.s. of eq. (\ref{second}). As expected $\lambda(T)$ is smaller than $m_G(T)$ below $T_d$ and therefore the 1RSB branch of solutions cannot be continued below $m_G(T)$ near $T_d$. However we see that line $\lambda(T)$ crosses the marginal line at a temperature $T_*=501227$. This confirms {\it a posteriori} the argument of the previous sections that in general the existence of $T_G$ implies the existence of $T_*$. For temperatures $T<T_*$ the 1RSB branch of solutions can be continued to values of the breaking point $m<m_G(T)$ by considering a FRSB ansatz.

According to what we said in subsection \ref{suboff} we expect that both the marginal condition {\it and} the condition $\lambda_{eff} \equiv \lambda/m<1$ are necessary in order for the 1RSB solution relevant for off-equilibrium dynamics. The thin solid line in fig. (\ref{figTm}) represents $\lambda_{eff}(T)$, we see that it starts from the value $\lambda_{eff}=\lambda=.743$ at $T_d$ (consistently with \cite{Ferrari12}) and increases up to $\lambda_{eff}=1$ at $T_*$. This determines a qualitative change in off-equilibrium dynamics at $T_*$ and implies that the marginal 1RSB solution must be abandoned below $T_*$ in order to describe off-equilibrium dynamics because $\lambda_{eff}>1$.  On the other hand below $T_*$ a continuous branch of FRSB solutions appears and the end-point of the branch is the natural candidate to describe off-equilibrium dynamics.

\begin{figure}[htb]
\begin{center}
\includegraphics[width=10cm]{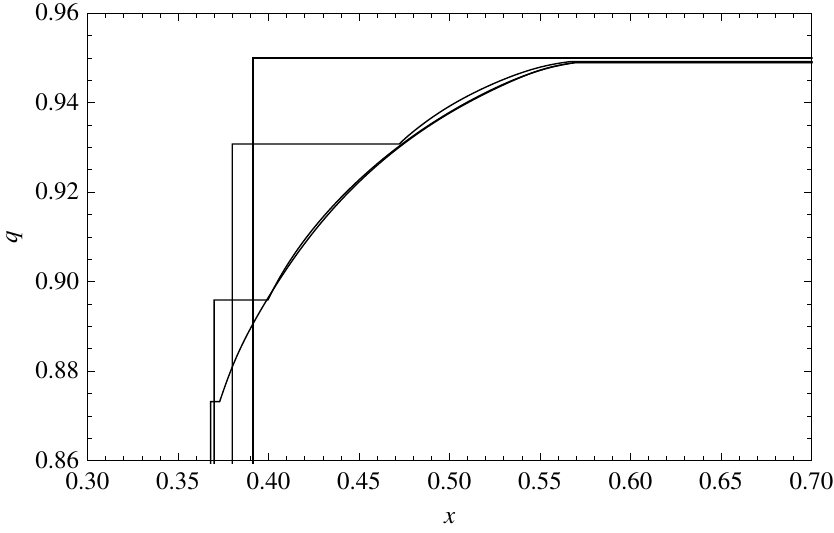}
%\put(-270,80){\rotatebox{90}{\it q}}
%\put(-120,-10){\it x}
\caption{The $q(x)$ of the Ising $p$-SG for $p=3$ for different values of the breaking point $m=.368,.37,.38,.3914$ at $T=.3<T_*$. The length of the first plateau decreases linearly to zero as $m$ approaches the end point $m_{end}=.3677$. $q(x)=0$ for $x<m$.}
\label{figqxX}
\end{center}\end{figure}
In Fig. (\ref{figqxX}) we plot $q(x)$ at $T=.3$ for various values of the breaking point $m$. At $m=m_G(.3)=.3914$ we have the marginal 1RSB solution. For smaller values of $m=.38,.37,.368$ the solution becomes FRSB with two plateaus. As expected according to section (\ref{onset}), for $m$ near $m_G$ the continuous region is concentrated at values of $x$ near $\lambda(T)=.57268$, and actually the starting point $x_P$ of the second plateau and the values of $q(x_P)$ do not change too much even at lower values. The end point $x_p$ of the first plateau instead decreases for $m<m_G$ until the end point $m=m_{end}=.3677$ where the first plateau has zero length. The FRSB solution cannot be continued to lower values of $m$ because we would have a negative plateau. As we can see in figure (\ref{figTm}) the line $m_{end}(T)$ (solid, gray) is almost a straight line connecting the point $(T_*,m_*)=(.501227,.61825)$ and the point $(0,0)$. 
Technically the numerical procedure used to solve the equation breaks down at $m_{end}$ and therefore its value was estimated by extrapolation, plotting parametrically the lengths of the first plateau $l_1$ as a function of $m$ and extrapolating $m$ to the point $l_1=0$. This procedure confirms that, as we saw in the previous section, the function $l_1(m)$ is regular near $m_{end}$  and also that the $q(x)$ at $x=m_{end}$ is regular.
\begin{figure}[htb]
\begin{center}
\includegraphics[width=10cm]{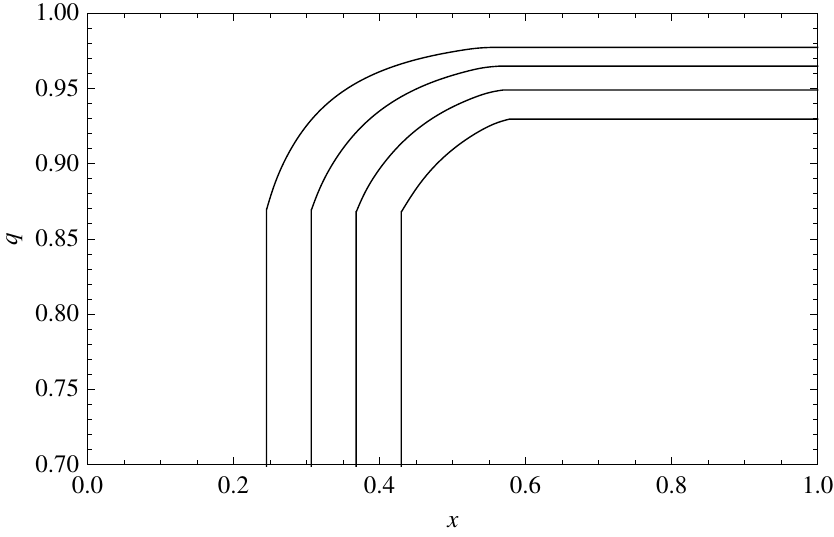}
%\put(-270,80){\rotatebox{90}{\it q}}
%\put(-120,-10){\it x}
\caption{The end-point solution $q(x)$ of the Ising $p$-SG for $p=3$ 
for different values of the temperature, from top to bottom $T=.2,.25,.3,.35$. $q(x)=0$ for $x<m_{end}(T)$}
\label{figqxT}
\end{center}\end{figure}
In figure (\ref{figqxT}) we plot the function $q(x)$ for $m=m_{end}(T)$ for $T=.35,.3,.25,.2$. They were actually obtained choosing a value of $m$ as close as possible to $m_{end}$. A computation down to zero temperature is feasible,  possibly by means of the methods of \cite{Sommers84}, but it goes beyond the scope of this work.

\begin{figure}[htb]
\begin{center}
\includegraphics[width=10cm]{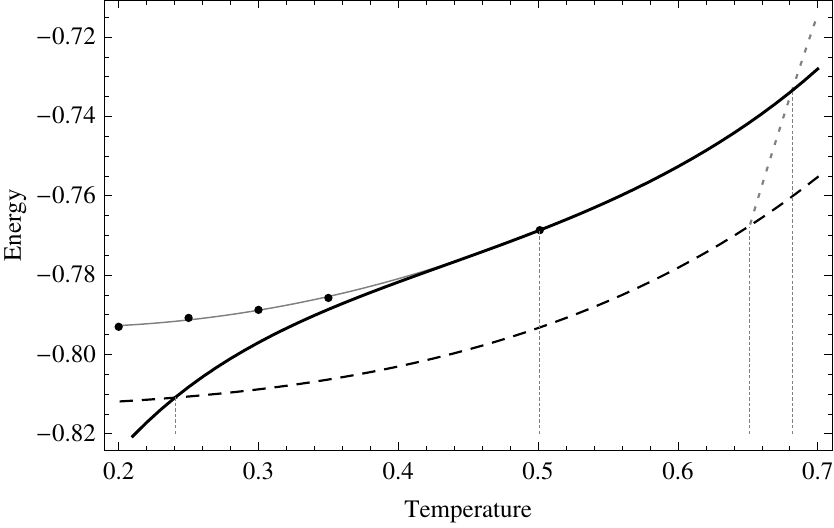}
%\put(-270,80){\rotatebox{90}{\it E}}
%\put(-130,-10){\it T}
\caption{Energy vs. Temperature plot of the various solutions of the Ising $p$-SG with $p=3$. The thin vertical lines represent the temperatures $T_G=0.24026$, $T_*=.501227$ $T_s=.651385$ and $T_d=.681598$. Dotted line: energy of the paramagnetic solution $E_{para}=-\beta/2$. Dashed line: energy of the static 1RSB solution $E_s(T)$ that gives the equilibrium energy between $T_G$ and $T_s$.
Solid thick line: energy $E_G(T)$ of the marginal 1RSB solution.
The points at $T=.2,.25,.3,.35,T_*=.501227$ are the values of the energy $E_{end}(T)$ of the end-point of the FRSB branch. 
Solid thin line: quadratic fit between the five points reported in the text. For any model the end-point energy $E_{end}(T)$ must be tangent to $E_G(T)$ at $T=T_*$.
}
\label{figenergie}
\end{center}\end{figure}

The energy of a given solution is given by: 
\beq
E=-{\beta\over 2}\left(1-\int_m^1 q^p(x)dx\right)
\eeq
In figure (\ref{figenergie}) we plot the energy of various solutions of the variational equations as a function of the temperature for $p=3$.
The dotted line is the energy $-\beta/2$ of the paramagnetic solution that gives the equilibrium value for $T>T_s=.651385$. The dashed line is instead the energy of the 1RSB solution that extremizes the free energy with respect to $m$ and that yields the equilibrium energy in the temperature range $(T_G,T_s)$. The solid line is the energy $E_G(T)$ of the marginal solution. The energy of the marginal solution coincides with the equilibrium energy at $T_d$ where equilibrium dynamics has the MCT-like dynamical singularity.
Between $T_d$ and $T_*$ the marginal solution is a natural candidate to describe off-equilibrium dynamics. Below $T_*$ the marginal solution is not consistent with off-equilibrium dynamics because $\lambda_{eff}>1$ and  
the natural candidate becomes instead the end point of the FRSB branch.
As it was done for $m_{end}(T)$ the energy of the end-point can be obtained plotting parametrically the energy as a function of $l_1$ (the length of the first plateau) and extrapolate to $l_1=0$. The procedure however is affected by large numerical errors that become larger both near $T_*$ and near zero temperature. In figure we plot the numerical estimates for four temperatures reported in the following table:
\begin{center}
  \begin{tabular}{ l | c | c | c | c }
    $T$ & \  .2 \  & \  .25 \  &  \ .3 \ & \  .35 \ \\ \hline
    $E_{end}$ \  &  \ -.7931 & \ -.7908 & \  -.7888 & \  -.7858  \\
    $m_{end}$ \ & .2444 & .3054 & .3677 & .4289 
  \end{tabular}
\end{center}
The above values for $m_{end}$ where used in order to draw the line $m_{end}(T)$ in fig. (\ref{figTm}) by interpolation.
The values for the energy together with the (much more precise) value $E_*=-.768700$ at $T_*=.501227$ are well fitted by the following quadratic form which is also plotted in figure (\ref{figenergie}):
\beq
E_{end}(T)= -.78829 -.06307 \times T + .20354 \times T^2
\label{fitene}
\eeq
The above simple fit should be only used for interpolation in the range of temperatures $(.2,.501227)$ and is certainly not accurate for lower temperatures where a more refined numerical analysis should be made.

It is interesting to consider the behavior of the FRSB solution near $T_*$.
If we go back to eqs. (\ref{plateaus}) we see that for temperatures $T=T_*+\Delta T$ near $T_*$ the quantity $(x_1-n)$ on the marginal solution  is $O(\Delta T)$. Continuing the marginal solution to lower values of the breaking point $m=m_G+\Delta m$ the replicon $m_1$ is proportional to $\Delta m$ and we have: $\delta q(m) \propto \Delta m/\Delta T$ and $\delta q(1) \propto \Delta m$.
Assuming that the $q(x)$ has a finite derivative, the size of the continuous  region $\Delta x$ separating the two plateaus grows linearly with their difference in height leading to $\Delta x \propto \Delta m/\Delta T$. The end point of the branch is located where $\Delta x$ becomes comparable to $x_1-n$ from which we obtain $\Delta m_{end}=O(\Delta T^2)$.  It is easily seen that also the energy has the same behavior meaning that {\it $m_{end}(T)$ and $E_{end}(T)$  are tangent to the corresponding Gardner lines at $T_*$}:
\beqa
m_{end}(T) & = & m_G(T)+O(T-T_*)^2
\\
E_{end}(T) & = & E_G(T)+O(T-T_*)^2
\eeqa
Note that the above result is model-independent and may be useful in situations where the actual solution of the FRSB equations is unfeasible. From fig. (\ref{figenergie}) we see that the fit (\ref{fitene}) of the $p=3$ model reproduces quite accurately this property of the true $E_{end}(T)$. 

We conclude this section with some technical remarks on the numerical solutions of the variational equations. Following \cite{Sommers84,Crisanti02} we have used an iterative procedure that involves discretization of the functions $P(x,y)$ and $\mu(x,y)$ on a two-dimensional grid $(x,y)$.
For fixed breaking point $m$ we start from an initial linearly increasing $q(x)$ defined between $m$ and  $1$ and evaluate the functions $P(x,y)$ and $\mu(x,y)$ by means of eqs. (\ref{eqP}) and (\ref{eqmu}), then a new value of $q(x)$ is obtained by means of eq. (\ref{qvar}) and the process is iterated.
For values of $m$ larger than $m_G(T)$ $q(x)$ converges to a constant corresponding to the 1RSB solution, while for $m<m_G(T)$ and $T<T_*$ a non-constant solution can be found down to values slightly larger than $m_{end}(T)$. Technically an important point is that some smoothing of the $q(x)$ must be applied at each iteration in order to avoid that it develops too higher derivatives making the use of the differential equations not appropriate.

A more subtle technical issue it that the derivative of the true solution has a discontinuity at the points $x_p$ and $x_P$ where the continuous part is joined with the plateaus.
As observed already in \cite{Crisanti02} the numerical solution tends to be rounded near these points due to the discretization. This effect can be removed if one has precise estimates of the location of $x_p$ and $x_P$ and solve the equations only in the region where $q(x)$ has non-zero derivative. In order to obtain such an estimate a rather complex procedure was suggested in \cite{Crisanti02} (see figs. (7,8,9) in that paper). Instead a direct estimate of the breaking points can be quickly obtained using eq. (\ref{second}) and we employed this method in order to update the position of the breaking points at each iteration.

\section{Numerical Simulations}
\label{numerics}

In order to validate the scenario put forward in section (\ref{suboff}) I have studied off-equilibrium dynamics by means numerical simulations. 
The results are quite interesting but not conclusive, further studies are needed in order to settle the issue. 
It turns out, in brief, that the off-equilibrium decay of the energy  $E(t)$ at $T=T_*$ can be fitted by a power law {\it but} with a limiting value $E(\infty)$ higher than $E_*$. This is consistent with the fact that {\it if} $E(\infty)=E_*$ dynamics must be slower than a power-law. However we cannot decide if the limiting value is actually $E_*$ or it is higher. The same phenomenon occurs for the remanent magnetization $m(t)$ with the important difference that in this case the standard expectation is that $m(\infty)=0$ while a power law extrapolation yields $m(\infty)>0$. Finally a parametric plot of the energy vs. the remanent magnetization supplemented with the assumption $m(\infty)=0$ yields a value $E(\infty)$ consistent with $E_*$ within the overall precision.

Numerical simulations of the fully connected Ising $p$-spin model of size $N$ requires $O(N^p)$ interactions and are therefore limited to relatively systems sizes. In order to overcome this problem I have considered systems with large connectivities and extrapolated to the infinite connectivity limit. 
In the simulations I considered a set of $N$ variable nodes (the Ising spins $s_i=\pm 1$) and a fixed number $\alpha N$ of $3$-spin factor nodes. Each factor node is connected randomly to three variable nodes $ijk$ and a quenched random coupling $J_{ijk}=\pm 1$ is assigned to it. The average connectivity of each site is thus $c \equiv 3 \alpha$. In order to compare systems at finite connectivities with the fully connected model one has to choose the temperature according to $\beta \equiv \beta' /\sqrt{2 c}$, where $\beta'$ is the target temperature in the fully connected model. Dynamics is standard Monte-Carlo starting from a random configuration. I measured the time decay of the energy and of the remanent magnetization, defined as the overlap between the initial configuration and the configuration at time $t$. 

\begin{figure}[htb]
\begin{center}
\includegraphics[width=10cm]{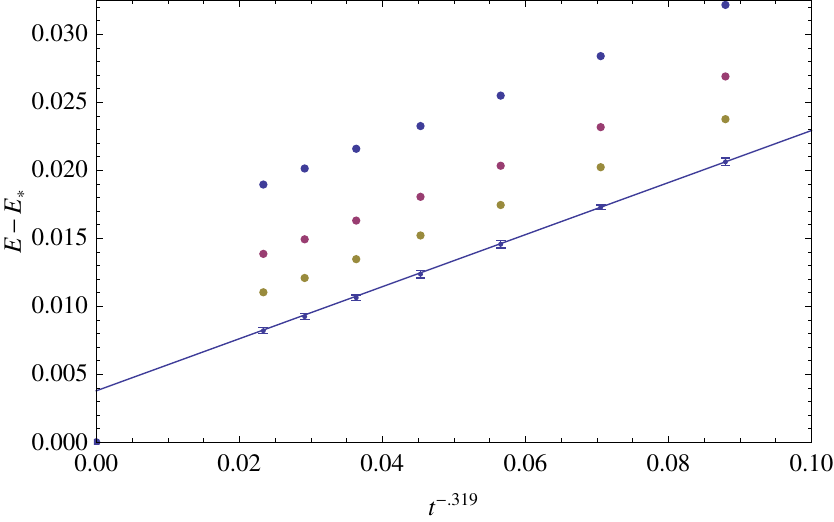}
%\put(-270,60){\rotatebox{90}{$E-E_*$}}
%\put(-130,-10){\it $t^{-.319}$}
\caption{Off-Equilibrium Dynamics at $T_*$ in $3$-spin Ising model. Plot of the energy minus the marginal energy $E_*$ vs. $t^{-.319}$ in MCS units. From Top to bottom we have $E_{24}-E_*$, $E_{45}-E_*$, $E_{90}-E_*$ and $E_{est}-E_*$ where $E_{est} \equiv 2 E_{90}-E_{45}$. 
Error bars are smaller than the points when not shown. 
The straight line is the three parameters fit $E_{est}-E_*=.0038+.191\, t^{-.319}$. 
The points correspond to $t=2^{k}$ with $k=11,\dots,17$ and the values of the energy are time averages over the corresponding time intervals and over ten runs. System size is $N=10^6$.
}
\label{figET}
\end{center}\end{figure}

In figure (\ref{figET}) we plot the decay of the energy as a function of the number of MCS steps for connectivities  $c=24,45,90$, system size $N=10^6$, $T=T_*=.501227$ at times $t=2^k$ with $k=11,\dots,17$. The data were obtained from  10 runs of $2^{17} MCS$ and a new random graph is generated at each run. The value of the energy at time  $t=2^k$ is an average over the time interval $(2^{k-1}),2^k$.  Assuming $O(1/c)$ corrections induced by the finite connectivity, an estimate for the infinite value limit is given by $E_{est}=2 E_{90}-E_{45}$.  Corrections to the estimate {\it in the considered time range} are negligible within the overall precision as was confirmed by an analysis at lower connectivity $c=24$. Data for $N=10^6/2$ (not shown) are superimposed (within the errors) with the corresponding data at $N=10^6$ and therefore we assume that we are sufficiently close to the thermodynamic limit.

The data are shifted vertically by an amount $E_*=-.768700$, according to the result of the previous section. We see that the estimated $E(t)$ is compatible with a power-law decay $1/t^a$ with an exponent $a=.319$ obtained from a three parameter fit $E_{est}(t)-E_*=.0038+.191\, t^{-.319}$. This leads to $E(\infty)-E_*=.0038$, {\it i.e.} the limiting value of the energy would be definitively larger than $E_*$. Note also that this deviation is significant also on the scale of fig. (\ref{figenergie}).

In a sense these results are compatible with the scenario we put forward in the previous sections. According to it the energy decays to $E_*$ slower than any power law at $T=T_*$. Therefore if we fit the data in a limited time window with a power law we would get a (wrong) estimate definitively larger than $E_*$.  
Nevertheless this is not a very strong evidence and we cannot rule out the fact that the scenario is wrong altogether, {\it i.e.} that the decay is really power law and the limiting value of the energy is definitively larger than $E_*$.

\begin{figure}[htb]
\begin{center}
\includegraphics[width=10cm]{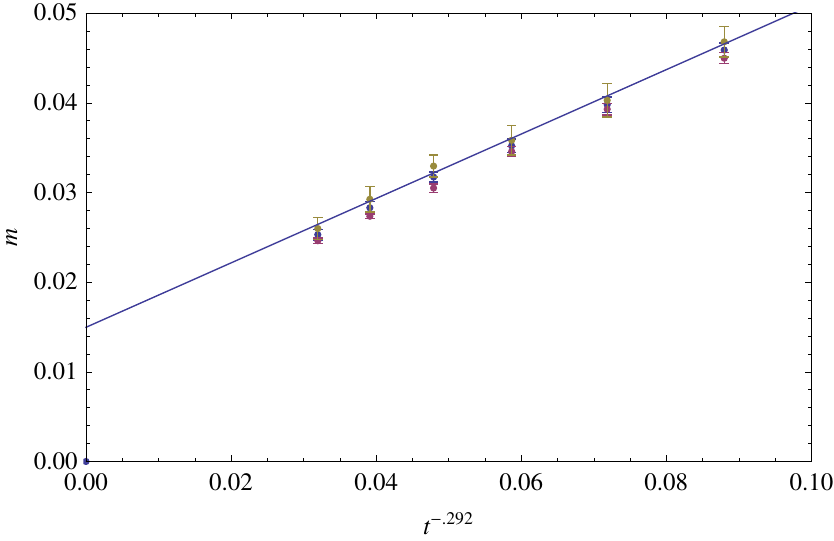}
%\put(-270,80){\rotatebox{90}{$m$}}
%\put(-130,-10){\it $t^{-.292}$}
\caption{Off-Equilibrium Dynamics at $T_*$ in $3$-spin Ising model. Plot of the remanent magnetization minus vs. $t^{-.292}$ in MCS units. From bottom to top we have $m_{45}$, $m_{90}$ and $m_{est}\equiv 2 \, m_{90}-m_{45}$.
The straight line is a three-parameters fit on the $m_{est}$ data $m(t)=.0150+.359\, t^{-.292}$. The points correspond to $t=2^{k}$ with $k=12,\dots,17$ and the values of the remanent magnetization are time averages over the corresponding time intervals and over ten runs. System size is $N=10^6$.
}
\label{figQT}
\end{center}\end{figure}

More insight comes from the study of the remanent magnetization.
In figure (\ref{figQT}) we plot $m_{45}$, $m_{90}$ and $m_{est}\equiv 2 \, m_{90}-m_{45}$  as a function of $t^{-.292}$ for the same runs of fig. (\ref{figET}).
As before the value at time $t=2^k$ is obtained as an average over the time interval $(2^{k-1}),2^k$ for each run. Once again   we see that  $m_{est}(t)$ is compatible with a power-law decay $1/t^a$ with an exponent $a=.292$ obtained from a three-parameters fit $m(t)=.0150+.359\, t^{-.292}$. Note however that the infinite limit of the remanent magnetization would be different from zero and, much as in the case of the energy, it seems that the difference, although small, is definitively larger than the overall error. The picture is similar to what we found for the energy except for the fact that the expectation that the remanent magnetization decay to zero is much more standard than the  expectation that the energy decays to $E_*$. Indeed it is related to ``weak long-term memory'' which is a key assumption within Cugliandolo-Kurchan theory \cite{Cugliandolo95}. 

The above results for the energy and remanent magnetization are compatible with the fact that they decay to their limiting values, respectively $E_*$ and zero, slower than any power law. 
One could just say that the time-scales explored are too small to display the asymptotic behavior. As a consequence one would expect that the data do not contain precise information on the limiting values. Surprisingly instead it turns out that if we plot  parametrically the energy vs. the remanent magnetization $(E(t),m(t))$ and {\it assume} that $m(\infty)=0$ the deviation of $E(\infty)-E_*$ is reduced to within the overall precision.
\begin{figure}[htb]
\begin{center}
\includegraphics[width=10cm]{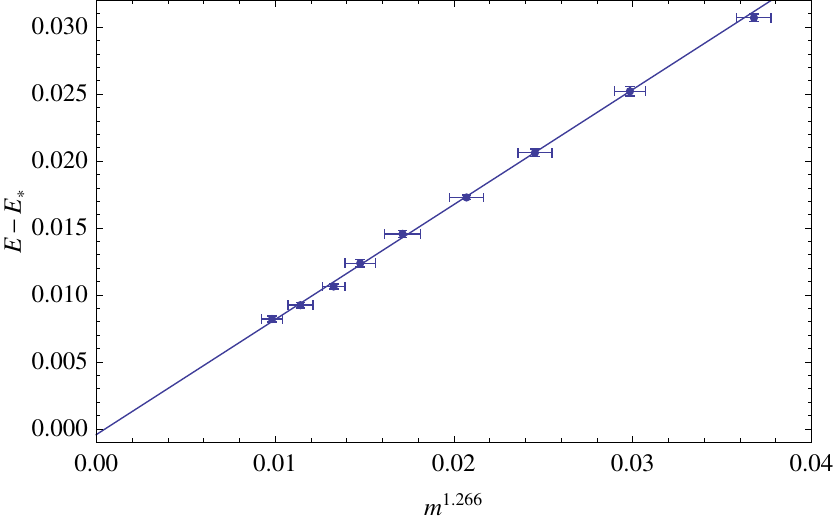}
%\put(-270,60){\rotatebox{90}{$E-E_*$}}
%\put(-130,-10){\it $m^{1.266}$}
\caption{Off-Equilibrium Dynamics at $T_*$ in $3$-spin Ising model. Parametric plot of the estimated energy $E_{est}$ minus $E_*$ vs. the estimated magnetization $m_{est}$ to the power $1.266$, datas as in the previous figures. 
The straight line is the three parameters fit $E_{est}-E_*=-.000428+.859 m_{est}^{1.266}$. 
}
\label{figEQ}
\end{center}\end{figure}
In figure (\ref{figEQ}) the estimated energy is plotted parametrically as a function of $m_{est}^{a}$, where the exponent $a=1.266$ is obtained from a three-parameters fit $E(t)-E_*=-.000428+0.859 m^{1.266}$. We see that while a power-law fit on $E(t)$ gives a deviation $E(\infty)-E_*=.0038$ with an error not compatible with zero, a parametric plot supplemented with $m(\infty)=0$ reduces the deviation to $E(\infty)-E_*=-.000428$ which is clearly compatible with zero within the errors.

\section{Conclusions}
\label{conclusions}
We have shown that near the dynamical transition temperature it is not possible to stabilize the  1RSB solution beyond the marginal point by making a FRSB ansatz. 
This may change at a temperature $T_*$ strictly lower than $T_d$ below which the 1RSB branch can be continued to a FRSB branch.
The existence of a $T_*$ temperature depends on the detail of model considered, but we showed that it certainly exists for models that display the so-called Gardner transition and in this case $T_G<T_*<T_d$.
The above results follows solely from the structure of the replicated Gibbs free energy near $T_d$ and therefore are quite general. They were indeed confirmed by the study of the Ising 3-spin model. They are also in agreement with recent results in the context of RSB theory for dense amorphous hard-spheres in high dimension, that also exhibit a Gardner transition as a function of the packing fraction \cite{Kurchan13}. 

The FRSB branch of solution below $T_*$ was studied analytically for the truncated model and it is characterized by a two plateau structure.
The branch ends where the length of the first plateau vanishes because analytical continuation to lower values $m<m_{end}(T)$ would require a plateau of negative length. These features have been confirmed in the context of the Ising $p$-spin with $p=3$ by numerical solution of the FRSB equations.
Note that the transition occurring at $T_*$ is not an ordinary 1RSB-FRSB transition, indeed $(T_*,m_*)$ is actually a critical point that marks the end of a line of ordinary 1RSB-FRSB transitions occurring on the line $m_G(T)$ for $T<T_*$.  

The results were discussed in connection with off-equilibrium dynamics within Cugliandolo-Kurchan theory. I considered a scenario where the RSB solution relevant for off-equilibrium dynamics is the 1RBS marginal solution in the range $(T_*,T_d)$  and it is the end-point of the FRSB branch for $T<T_*$.
Remarkably under these assumptions it can be argued that $T_*$ marks a qualitative change in off-equilibrium dynamics in the sense that the effective parameter exponent $\lambda_{eff}$ goes to one at $T_*$ and as a consequence the decay of various dynamical quantities changes from power-law to logarithmic. This suggests that the critical point $(m_*,T_*)$ is the off-equilibrium analog of the so-called $A_3$  singularity in equilibrium MCT which is also characterized by logarithmic decays \cite{Gotze89}. These peculiar dynamical features could be relevant in the context of aging numerical experiments in randomly packed soft sphere that have been reported recently \cite{Okamura13}.

Numerical simulations are consistent with the above scenario but further studies are needed in order to assess its validity. One possible route is to reconsider models on Bethe lattices \cite{Montanari04} supplementing the analysis of the data with the computation of $T_*$ in these models.  Besides numerical simulations one could also solve numerically the off-equilibrium dynamical equations in the appropriate spherical models \cite{Crisanti07}, possibly by means of adaptive algorithms \cite{Kim01,Lefevre06}.

{\em Acknowledgments.} ~~ This work originated from some stimulating discussions initiated by F. Zamponi and joined by F. Krzakala, J. Kurchan, G. Parisi and F. Ricci-Tersenghi, it is a pleasure to thank  them.
The European Research Council has provided financial support through ERC grant agreement no. 247328.


\begin{thebibliography}{99}



\bibitem{Kirkpatrick1987} T. R. Kirkpatrick, D. Thirumalai, Phys. Rev. B36, (1987) 5388

\bibitem{Kirkpatrick1989} T. R. Kirkpatrick, D. Thirumalai, J. Phys. A. vol. 22, L149 (1989)

\bibitem{Crisanti92} A. Crisanti, H. J. Sommers, Z. Phys. B 87 (1992) 341. 

\bibitem{Crisanti93} A. Crisanti, H. Horner and H. J. Sommers, Z. Phys. B92 (1993) 257. 

\bibitem{Cugliandolo93} L. F. Cugliandolo and J. Kurchan, Phys. Rev. Lett. 71 (1993) 173.

\bibitem{Monasson95} R. Monasson, Phys. Rev. Lett. 75 (1995) 2847

\bibitem{Franz97} S. Franz and G. Parisi, Phys. Rev. Lett. 79 (1997) 2486. 

\bibitem{Gotze09} W. G{\"o}tze, Complex Dynamics of Glass-Forming Liquids: A
  Mode-Coupling Theory, OUP (Oxford, UK), 2009.

\bibitem{ParisiRizzo13} G. Parisi and T. Rizzo, Phys. Rev. E 87, 012101 (2013).  

\bibitem{Cugliandolo96} L. F. Cugliandolo and P. Le Doussal, Phys. Rev. E 53, 1525 (1996).

\bibitem{Almeida78} J R L de Almeida and D J Thouless, J. Phys. A: Math. Gen. 11, 983 (1978).

\bibitem{Marinari94} E. Marinari, G. Parisi and F. Ritort, J. Phys. A: Math. Gen. 27, 7647 (1994).

\bibitem{Montanari03} A. Montanari and F. Ricci-Tersenghi, Eur. Phys. J. B 33, 339 (2003). 

\bibitem{Crisanti05} A. Crisanti, L. Leuzzi and T. Rizzo, Phys. Rev. B 71, 094202 (2005). 

\bibitem{Cugliandolo95} L. Cugliandolo and J. Kurchan, Phil. Mag. B, 71, 501 (1995).

\bibitem{Tonosaki07} Y. Tonosaki, K. Takeda, and Y. Kabashima, Phys. Rev. B 75, 094405 (2007). 

\bibitem{Gardner85} E. Gardner, Nucl. Phys. B 257, 747 (1985). 

\bibitem{Franz11} S. Franz, G. Parisi, F. Ricci-Tersenghi and T. Rizzo, Eur. Phys. J. E. (2011) 34: 102.

\bibitem{Parisi79} G. Parisi, Phys. Rev. Lett. 43, 1754–1756 (1979). 

\bibitem{DeDominicis98} C. De Dominicis, T. Temesvari and I. Kondor, J. Phys. IV France 08 (1998) Pr6-13 (arXiv:cond-mat/9802166).

\bibitem{Bouchaud98} J.-P. Bouchaud, L. Cugliandolo, J. Kurchan and M. Mezard, in `Spin-glasses and random fields', A. P. Young Ed. (World Scientific) 1998. 

\bibitem{Caltagirone13} F. Caltagirone, G. Parisi and T. Rizzo, Phys. Rev. E 87, 032134 (2013).

\bibitem{Gotze89} W. G{\"o}tze and L. Sj{\"o}gren, J. Phys.: Cond. Matt. 1, 4203 (1989).

\bibitem{Franz95} S. Franz, E. Marinari and G. Parisi, J. Phys. A: Math. Gen. 28 5437 (1995).

\bibitem{Kim01}  B. Kim and A. Latz, Europhys. Lett. 53, 660 (2001).

\bibitem{Lefevre06}  A. Andreanov and A. Lefevre, Europhys. Lett. 76, 919 (2006).

\bibitem{Montanari04} A. Montanari and F. Ricci-Tersenghi, Phys. Rev. B 70, 134406 (2004).

\bibitem{Gross85} D. Gross, I. Kanter, and H. Sompolinsky, Phys. Rev. Lett. 55, 304 (1985).

\bibitem{Temesvari02} T. Temesvari, C. De Dominicis and I. R. Pimentel, Eur. Phys. J. B 25, 361-372 (2002). 

\bibitem{Parisi80} G. Parisi, J. Phys. A: Math. Gen. 13 1887, (1980).

\bibitem{Kirkpatrick87} T. Kirkpatrick and D. Thirumalai, Phys. Rev. Lett. 58,
2091 (1987).

\bibitem{Ferrari12} U. Ferrari, L. Leuzzi, G. Parisi and T. Rizzo, Phys. Rev. B 86, 014204 (2012). 

\bibitem{Sommers84} H. J. Sommers, W. Dupont, J. Phys. C 17 (1984) 5785-5793.

\bibitem{Sommers85} H.-J. Sommers, J. Physique Lett. 46, L-779 (1985).

\bibitem{Crisanti02} A. Crisanti and T. Rizzo, Phys. Rev. E 65, 046137 (2002).

\bibitem{Kurchan13} J. Kurchan, G. Parisi, F. Zamponi, J. Stat. Mech. (2012) P10012,  J. Kurchan, G. Parisi, P. Urbani and F. Zamponi, 	arXiv:1303.1028 and private communication.

\bibitem{Okamura13} S. Okamura and H. Yoshino, arxiv:1306.2777

\bibitem{Crisanti07} A. Crisanti and L. Leuzzi, Phys. Rev. B 76, 184417 (2007)

 
\end{thebibliography}
\end{document}